\documentclass[conference]{IEEEtran}
\IEEEoverridecommandlockouts
\usepackage{cite}
\usepackage{amsmath,amssymb,amsfonts}
\usepackage{algorithmic}
\usepackage{graphicx}
\usepackage{textcomp}
\usepackage{xcolor}
\usepackage{booktabs}
\usepackage{threeparttable}
\usepackage{multirow}
\usepackage{marvosym}
\def\BibTeX{{\rm B\kern-.05em{\sc i\kern-.025em b}\kern-.08em
		T\kern-.1667em\lower.7ex\hbox{E}\kern-.125emX}}

\begin{document}
\title{Uniformly Accelerated Motion Model\\ for Inter Prediction\vspace{-0.3em}}

\author{
Zhuoyuan Li,
Yao Li,
Chuanbo Tang,
Li Li,
Dong Liu\textsuperscript{\Letter},
Feng Wu\\[0.2em]

\textit{University of Science and Technology of China (USTC),}\vspace{-0.1em}

\textit{Hefei, Anhui 230027, China}\\[0.1em]

\{zhuoyuanli, mrliyao, cbtang\}@mail.ustc.edu.cn, \{lil1, dongeliu,  fengwu\}@ustc.edu.cn\vspace{-1em}
\thanks{
	\Letter: Corresponding author.}
}
\maketitle
\begin{abstract}
Inter prediction is a key technology to reduce the temporal redundancy in video coding. In natural videos, there are usually multiple moving objects with variable velocity, resulting in complex motion fields that are difficult to represent compactly. In Versatile Video Coding (VVC), existing inter prediction methods usually assume uniform speed motion between consecutive frames and use the linear models for motion estimation (ME) and motion compensation (MC), which may not well handle the complex motion fields in the real world. To address these issues, we introduce a \underline{u}niformly \underline{a}ccelerated \underline{m}otion \underline{m}odel (\underline{UAMM}) to exploit motion-related elements (velocity, acceleration) of moving objects between the video frames, and further combine them to assist the inter prediction methods to handle the variable motion in the temporal domain. Specifically, first, the theory of UAMM is mentioned. Second, based on that, we propose the UAMM-based parameter derivation and extrapolation schemes in the coding process. Third, we integrate the UAMM into existing inter prediction modes (Merge, MMVD, CIIP) to achieve higher prediction accuracy. The proposed method is implemented into the VVC reference software, VTM version 12.0. Experimental results show that the proposed method achieves up to 0.38\% and on average 0.13\% BD-rate reduction compared to the VTM anchor, under the Low-delay P configuration, with a slight increase of time complexity on the encoding/decoding side.
\end{abstract}

\vspace{0.3em}
\begin{IEEEkeywords}
Inter prediction, motion estimation (ME), motion compensation (MC), VVC, uniformly accelerated motion model (UAMM).
\end{IEEEkeywords}

\section{Introduction}
Block-based hybrid coding framework has been widely
adopted in the modern video coding standards\cite{sullivan2012overview, bross2021overview}. To reduce the temporal redundancy in sequential frames, inter prediction makes a major contribution to these standards and plays a key role in the hybrid video coding scheme \cite{chien2021motion, yang2021subblock}. Block-based inter predictive process is usually performed on the assumption that the motion between consecutive frames tends to be uniform, and uses the linear model for motion estimation (ME) and motion compensation (MC) to generate the prediction of coding block. However, in natural videos, there are usually multiple moving objects with variable velocity and irregular deformation, leading to complex motion fields. To tackle the efficient representation of complex motion, many  motion modeling methods have been intensively studied in the development of video coding standards. These methods can be classified into two categories: spatial/temporal motion modeling-based methods (SMM/TMM).

\textit{\textbf{Spatial motion modeling.}} In Versatile Video Coding (VVC), most of the inter prediction modes are designed based on SMM. In the spatial domain, the motion information of previous encoded/decoded blocks located around the current coding block is most relevant to the motion information of the current coding block. Based on this assumption, translational-based \cite{chien2021motion} and deformable-based \cite{yang2021subblock, li2017efficient, zhang2018improved, zhang2016merge} spatial motion modeling methods are widely studied. HEVC considers the translation (uniform speed) motion  model in inter prediction, such as advanced motion vector prediction (AMVP) and Merge. Based on these two basic approaches, in VVC, MMVD, CIIP, HMVP \cite{li2019history}, GPM\cite{gao2020geometric}, etc are gradually adopted as the extension modes, and improve the spatial-motion-information reference ability of coding block. In addition, the deformable motion model \cite{li2017efficient, zhang2018improved, zhang2016merge} is tentatively studied for higher complex motion modeling (such as rotation and zooming) in HEVC, and widely adopted in VVC and integrated with existing inter modes to improve the flexibility of motion information description effectively. Although the existing SMM-based methods can finely characterize the complex motion, the limitation exists that the SMM only considers the spatial correlation and ignores the variable temporal correlation of motion.

\begin{figure}[b]
	\centering
	\vspace{-1em}
	\includegraphics[width=68mm]{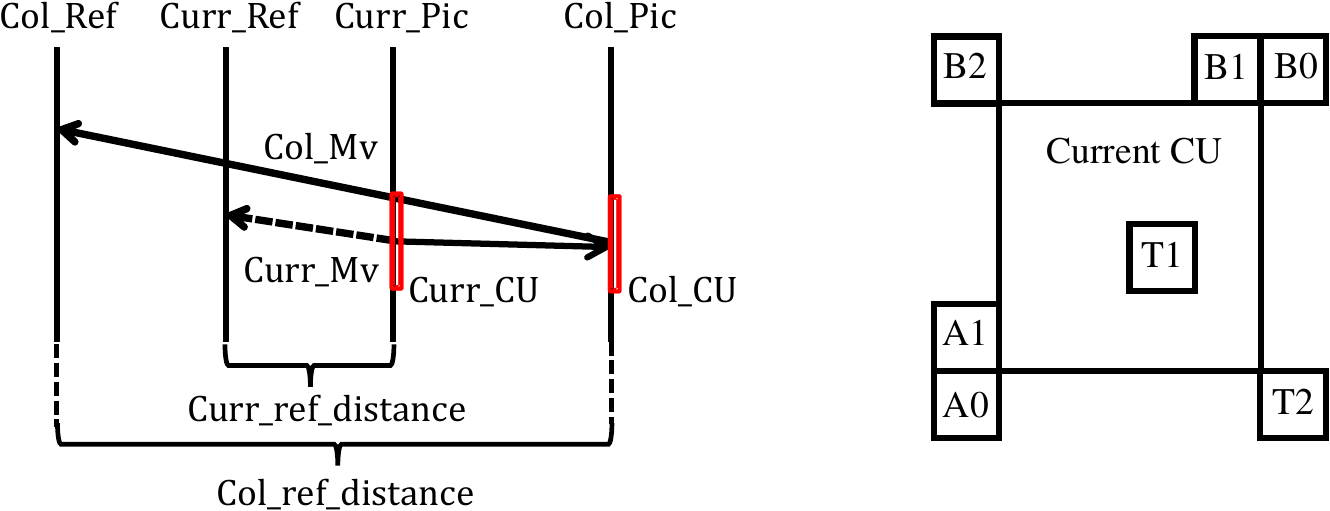}
	\begin{equation*}
		\scriptsize\label{qua}
	\begin{aligned}
		Curr\_Mv=\frac{Curr\_ref\_distance}{Col\_ref\_distance} \times Col\_Mv
	\end{aligned}
	\end{equation*}\vspace{-0.8em}
	\vspace{-1em}
	\caption{Illustration of temporal motion vector prediction (TMVP) in VVC.}
	\label{fig:arch}
\end{figure}

\begin{figure*}
	\centering
	\vspace{-2.5em}
	\includegraphics[width=135mm]{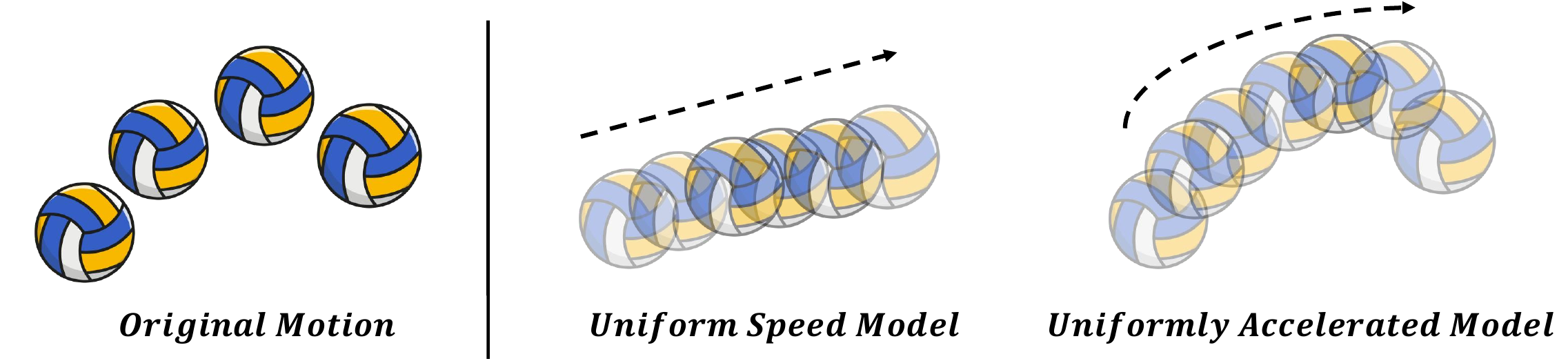}
	\vspace{-0.8em}
	\caption{Exploiting the uniformly accelerated model for temporal-domain motion modeling. Here we illustrate the different motion models. For the uniform speed model, the object motion can only be modeled on the linear trajectory (straight line). Compared to the uniform speed model, the uniformly accelerated model is used to model the motion with the velocity and acceleration elements in the $x$ and $y$ directions. The different accelerations of $x$ and $y$ directions enable it to model the motion on the nonlinear trajectory (curve), which makes the uniformly accelerated model enable the stronger representation ability.}
	\label{fig:introduction}
\end{figure*}

\begin{figure*}
	\centering
	\vspace{-0.3em}
	\includegraphics[height=35mm]{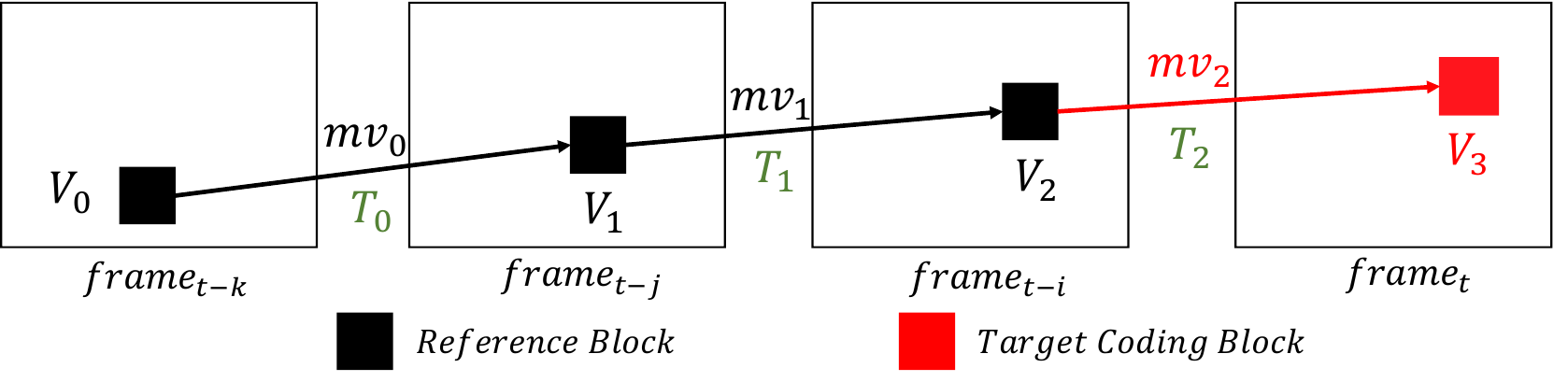}
	\vspace{-0.9em}
	\caption{Illustration of the motion trajectory between the coding (to-be-coded) block and reference (reconstructed) blocks in previous reference frames. The coding block and its motion information are colorized in red, the reference blocks and their motion information are colorized in black.}
	\vspace{-1.5em}
	\label{fig:intro}
\end{figure*}

\textit{\textbf{Temporal motion modeling.}} In the consecutive video frames, the motion propagates with the strong temporal correlation. To model the motion in the temporal domain, in HEVC, \textit{temporal motion vector prediction (TMVP)} \cite{McCann2013} has been adopted to derivate the temporal motion vector candidate by directly scaling the motion vector stored at adjacent location (T1, T2 in Fig.1) in the col-located picture to the target reference picture. Furthermore, in VVC, TMVP is integrated into more inter modes, and also extended to \textit{sub-block-based temporal motion vector prediction (SbtMvp)} \cite{yang2021subblock} to improve the accuracy of temporal motion vector prediction at the fine-grained block level. Although these two kinds of TMM-based methods bring the temporal motion vector prediction in inter modes, the performance is lower than the SMM-based method due to their simple temporal modeling strategy. 

Based on the limitations of these motion modeling methods, in this paper, we explore the high-order temporal motion model to achieve variable motion-aware temporal motion modeling in video coding scenes. Inspired by \cite{xu2019quadratic, liu2020enhanced, yang2022advancing}, we introduce the uniformly accelerated motion model (Fig.~\ref{fig:introduction}) to adapt to more complex motion situations. Rather than the uniform speed (linear) motion model, the uniformly accelerated (high-order) motion model can exploit the motion-related elements (velocity, acceleration) of moving objects, and approximate the complex motion with variable velocity and nonlinear trajectory based on a quadratic function. Specifically, we have made the following contributions that will be detailed in this paper:
\begin{itemize}
	\vspace{-0.1em}
	\item To the best of our knowledge, we are the first to introduce the uniformly accelerated motion model (UAMM) to improve the performance of inter prediction.
	\item To adapt to the video coding framework, we formulate the usage of UAMM, and propose the UAMM-based parameter derivation and extrapolation schemes in the predictive process.
	\item We integrate the uniformly accelerated temporal motion model into existing inter modes (Merge, MMVD, CIIP) of VVC to achieve higher prediction accuracy.
\end{itemize}

\section{Proposed Method}
\subsection{Theory of Uniformly Accelerated Motion Model}
In the real world, the motion of objects can be formulated as the integration of velocity over time, where the velocity can be expressed as the combination of initial velocity and the integral of acceleration over time:
\vspace{-0.3em}
\begin{equation}\label{x_phy}
	\boldsymbol{x}_{\boldsymbol{t}}=\int_0^t{\boldsymbol{v}\left( t \right) dt},
\end{equation}
\begin{equation}\label{v_phy}
	\boldsymbol{v}\left( t \right) =\boldsymbol{v}_0+\int_0^t{\boldsymbol{a}\left( t \right) dt},
\end{equation}
where $\boldsymbol{x}_{\boldsymbol{t}}$, $\boldsymbol{v} \left( t \right)$, $\boldsymbol{a} \left( t \right)$ denote the object's motion vector (Mv), velocity and acceleration at the $t$ moment, respectively. While most of the motion variations in the filming scene are not excessively dramatic, the above motion model can be degraded to a uniformly accelerated motion model with the parameter of an initial velocity  $\boldsymbol{v}_{0}$ and a constant acceleration $\boldsymbol{a}$:
\vspace{-0.1em}
\begin{equation}\label{x_qua}
	\boldsymbol{x}_{\boldsymbol{t}}=\boldsymbol{v}_0\cdot t+\frac{1}{2}\cdot \boldsymbol{a}\cdot t^2,
\end{equation}
\begin{equation}\label{v_qua}
	\boldsymbol{v}\left( t \right) =\boldsymbol{v}_0+\boldsymbol{a}\cdot t,
\end{equation}

\subsection{Uniformly Accelerated Motion Model-based Parameter Derivation and Extrapolation}

Based on the theory of UAMM, here we detail the combination of UAMM and the coding (prediction) process, and propose the derivation scheme of UAMM-related parameters of reconstructed (reference) blocks and the extrapolation scheme of to-be-predicted motion information of the coding block.

\subsubsection{Derivation of the UAMM-related parameters} To fit the motion trajectory between the coding (to-be-coded) block and reference blocks in previous reference frames, as shown in Fig.~\ref{fig:intro}\,, the motion information of two reference blocks ($\boldsymbol{mv}_0$, $\boldsymbol{mv}_1$, $\boldsymbol{T}_0$, $\boldsymbol{T}_1$) is used to derive the parameters of the uniformly accelerated motion model in the temporal domain. 

\begin{figure*}
	\centering
	\vspace{-2.5em}
	\hspace{-13em}\includegraphics[height=59mm]{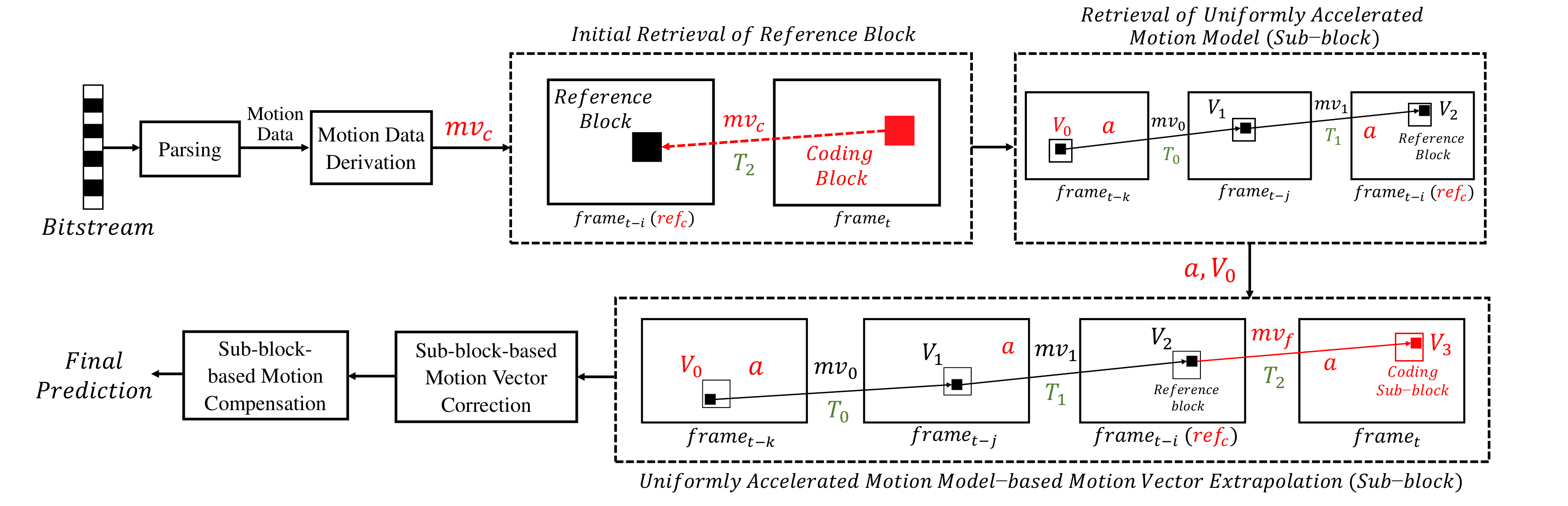}\hspace{-13em}
	\vspace{-1.3em}
	\caption{Illustration of the proposed uniformly accelerated motion model-integrated inter prediction (UAMM) framework from the perspective of decoding. Boxes represent the sub-modules of UAMM, and arrows indicate the direction of the data flow. The coding block and its motion information are colorized in red, the reference (reconstructed) blocks and their motion information are colorized in black.}
	\vspace{-1.8em}
	\label{fig:framework}
\end{figure*}

Combining the formulas (\ref{x_qua}), (\ref{v_qua}), the motion information of the reconstructed block in the third reference frame ($frame_{t-k}$) is served as the initial motion point of motion trajectory, and the correlation of motion information of reconstructed blocks can be constructed as the following equations:\vspace{-0.1em}
\begin{equation}\label{mv_0}\vspace{-0.1em}
	\boldsymbol{mv}_0=\boldsymbol{v}_0\cdot \boldsymbol{T}_0+\frac{1}{2}\cdot \boldsymbol{a}\cdot \boldsymbol{T}_{0}^{2},
\end{equation}\vspace{-0.3em}
\begin{equation}\label{mv_1}
	\boldsymbol{mv}_1=\boldsymbol{v}_1\cdot \boldsymbol{T}_1+\frac{1}{2}\cdot \boldsymbol{a}\cdot \boldsymbol{T}_{1}^{2},
\end{equation}\vspace{-0.5em}
\begin{equation}\label{v_1}
        \boldsymbol{v}_1=\boldsymbol{v}_0+\boldsymbol{a}\cdot \boldsymbol{T}_{0},
\end{equation}
where the $\boldsymbol{mv}_0$ and $\boldsymbol{mv}_1$ represent the motion vector of reconstructed blocks from $frame_{t-k}$ to $frame_{t-j}$ and $frame_{t-j}$ to $frame_{t-i}$, respectively, and $\boldsymbol{T}_{0}$, $\boldsymbol{T}_{1}$ denote the time interval between these three reference frames, as illustrated in Fig.~\ref{fig:intro}\,. Note that the time interval between two adjacent frames is a constant in the video, $\boldsymbol{T}$ is directly set as the distance of frame (POC) number. According to the (\ref{mv_0}), (\ref{mv_1}), (\ref{v_1}), the UAMM-related parameters of reconstructed blocks (initial velocity $\boldsymbol{v}_0$, acceleration $\boldsymbol{a}$) can be solved as follows:

\begin{equation}\label{solve_a}
        \boldsymbol{a}=\frac{2\left(\boldsymbol{mv}_1\cdot \boldsymbol{T}_0-\boldsymbol{mv}_0\cdot \boldsymbol{T}_1 \right)}{\boldsymbol{T}_0\boldsymbol{T}_1\left( \boldsymbol{T}_0+\boldsymbol{T}_1 \right)},
\end{equation}

\vspace{-0.5em}
\begin{equation}\label{solve_v_0}
        \boldsymbol{v}_0=\frac{\boldsymbol{mv}_0-\frac{\boldsymbol{a}\cdot \boldsymbol{T}_{0}^{2}}{2}}{\boldsymbol{T}_0}.
\end{equation}

In addition, if the acceleration ($\boldsymbol{a}$) or initial velocity ($\boldsymbol{v}_0$) is solved for 0, the UAMM degrades into a linear or constant model.

\subsubsection{Extrapolation of to-be-predicted motion information}
Based on the derivation process, the UAMM-related parameters of each reconstructed (reference) block can be solved and stored for the subsequent prediction process of coding (to-be-coded) block. For the extrapolation of to-be-predicted motion information, we assume that the motion trajectory is obtained with the proper indication, and introduce how to extrapolate the to-be-predicted motion information of the coding block. 

To solve the motion information of coding block in UAMM, as shown in Fig.~\ref{fig:intro}\,, the motion trajectory from the $frame_{t-k}$ to $frame_{t}$ is taken as a complete motion situation, and the stored acceleration and initial velocity of reconstructed (reference) block are shared to the subsequent coding block. The motion information of the current coding block can be solved as follows:\vspace{-0.5em}

\begin{equation}\label{mv_2_simple}
   \boldsymbol{mv}_2=\boldsymbol{v}_2\cdot \boldsymbol{T}_2+\frac{1}{2}\cdot \boldsymbol{a}\cdot \boldsymbol{T}_{2}^{2},
\end{equation}
and further can be written as the follows by decomposing $\boldsymbol{v}_2$ into $\boldsymbol{v}_0$, $\boldsymbol{a}$ and time intervals $\boldsymbol{T}_{0}$, $\boldsymbol{T}_{1}$:\vspace{-0.5em}
\vspace{-0.1em}
\begin{equation}\label{mv_2}
   \boldsymbol{mv}_2=\boldsymbol{v}_0\cdot \boldsymbol{T}_2+\boldsymbol{a}\cdot \boldsymbol{T}_2\left(\boldsymbol{T}_0+\boldsymbol{T}_1 \right) +\frac{1}{2}\cdot \boldsymbol{a}\cdot \boldsymbol{T}_{2}^{2}.
\end{equation}
Substituting the solved UAMM-related parameters of the formula (\ref{solve_a}) and (\ref{solve_v_0}) into (\ref{mv_2}), $\boldsymbol{mv}_2$ can be extrapolated, which is used to retrieve the reference block of the current coding block. \textit{Note that the parameters are all represented in vector form, and both the horizontal component and the vertical component can be derived simultaneously.}

\vspace{-0.3em}
\subsection{Integration of Uniformly Accelerated Motion Model and Inter Modes in VVC}
Based on the above calculation scheme of UAMM-related parameters, here we detail how to integrate the UAMM into the existing inter modes of VVC \cite{chien2021motion, yang2021subblock}. The whole prediction process is shown in Fig.~\ref{fig:framework} from the perspective of decoding. 

For the decoding process, first, the motion data of the current coding block is obtained by parsing the bitstream. Concerning the UAMM-related motion data for the coding block, our proposed method is integrated into the existing Merge, MMVD, and CIIP inter modes, and follows the decoding process of these three modes. Second, with the decoded motion information ($mv_c$, $ref_c$) obtained by indexing the motion vector candidate, the $mv_c$ and $ref_c$ are used to retrieve the reference block (ending point of motion trajectory) in reference (reconstructed) frames. Furthermore, the stored UAMM-related parameters ($\boldsymbol{v}_0$, $\boldsymbol{a}$) of each sub-block of reference block are retrieved and inherited for each sub-block of coding block. Note that the UAMM-related parameters of reference block are derivated in the $4\times4$ unit granularity and cached into the buffer of motion information, when the coding frame is reconstructed. Third, due to the difference of parameters between the sub-blocks of coding block, the sub-block-level UAMM-based motion vector extrapolation is performed to derive the motion information of each sub-block, respectively. Fourth, with the sub-block-level Mvs obtained, the motion vector correction is used to correct and refine these Mvs in the customized rules, including the constraint between the sub-block Mvs and initial Mv, the consistency of sub-block Mvs. Finally, the sub-block-based motion compensation is performed to generate the final prediction. 

For the encoding process, compared to the decoding process, only the process of rate-distortion-optimization (RDO) is increased. For the integration of each mode and UAMM, the UAMM-based motion vector refinement is used as a plug-and-play module after the process of motion vector derivation of each inter mode, without the transmission of any extra signal.

\begin{table}
	\renewcommand\arraystretch{1.3}
	\centering
	\fontsize{8.3pt}{9pt}\selectfont
	\vspace{-2em}
	\caption{Coding Performance and Relative Complexity of UAMM-integrated Inter Prediction Framework \\ Based  on VTM-12.0 under LDP Configuration}
	\label{tab:CTC}
	\setlength{\tabcolsep}{1.3mm}
	\vspace{-0.6em}
	\begin{tabular}{ccccccc}
		\cline{1-6}
		\multirow{2}{*}{\textbf{Class}}                                                        & \textbf{Sequence}        &           & \multicolumn{3}{c}{\textbf{Low-delay P}}               &           \\ \cline{2-2} \cline{4-6}
		& Name                     &           & Y                & U                & V                &           \\ \cline{1-2} \cline{4-6}
		\multirow{5}{*}{\textbf{\begin{tabular}[c]{@{}c@{}}ClassB\\ (1920x1080)\end{tabular}}} & \textit{MarketPlace}     &           & -0.11\%           & 0.43\%           & 0.47\%          &           \\
		& \textit{RitualDance}     &           & 0.01\%          & 0.10\%          & -0.04\%          &           \\
		& \textit{Cactus}          &           & -0.10\%          & -0.27\%          & -0.02\%          &           \\
		& \textit{BasketballDrive} &           & -0.12\%           & 0.05\%          & -0.03\%           &           \\
		& \textit{BQTerrace}       &           & 0.03\%          & 0.01\%          & -0.04\%          &           \\
		\cline{1-2} \cline{4-6}
		\multirow{4}{*}{\textbf{\begin{tabular}[c]{@{}c@{}}ClassC\\ (832x480)\end{tabular}}}   & \textit{BasketballDrill} &           & -0.16\%          & 0.22\%          & 0.02\%          &           \\
		& \textit{BQMall}          &           & -0.02\%          & -0.09\%           & -0.20\%           &           \\
		& \textit{PartyScene}      &           & -0.05\%          & -0.28\%          & 0.24\%          &           \\
		& \textit{RaceHorsesC}     &           & 0.06\%          & -0.46\%           & 0.70\%          &           \\
		\cline{1-2} \cline{4-6}
		\multirow{4}{*}{\textbf{\begin{tabular}[c]{@{}c@{}}ClassD\\ (416x240)\end{tabular}}}   & \textit{BasketballPass}  &           & -0.16\%          & 0.44\%          & -0.85\%           &           \\
		& \textit{BQSquare}        & \textit{} & -0.14\%          & -0.54\%          & -0.53\%          &           \\
		& \textit{BlowingBubbles}  & \textit{} & -0.15\%          & -0.82\%          & -0.12\%          &           \\
		& \textit{RaceHorses}      & \textit{} & -0.28\%          & 0.34\%          & -1.02\%          &           \\
		\cline{1-2} \cline{4-6}
		\multirow{3}{*}{\textbf{\begin{tabular}[c]{@{}c@{}}ClassE\\ (1280x720)\end{tabular}}}  & \textit{FourPeople}      &           & -0.35\%
		& -0.49\%
		& -0.43\%          &           \\
		& \textit{Johnny}          & \textit{} & -0.38\%          & 0.29\%          & -1.23\%          &           \\
		& \textit{KristenAndSara}  & \textit{} & -0.08\%
		& 0.27\%          & 0.14\%          &           \\
		\cline{1-2} \cline{4-6}
		\multicolumn{2}{c}{\textbf{Overall}}                                                                              & \textbf{} & \textbf{-0.13\%} & \textbf{-0.05\%} & \textbf{-0.17\%} & \textbf{} \\ \cline{1-6}
		\multicolumn{2}{c}{EncT}                                                                                          &           & \multicolumn{3}{c}{101\%}                              &           \\ \cline{1-6}
		\multicolumn{2}{c}{DecT}                                                                                          &           & \multicolumn{3}{c}{102\%}                              &           \\ \cline{1-6}
	\end{tabular}
	\vspace{-1.6em}
\end{table}

\vspace{-0.2em}
\section{Experimental Results}
\vspace{-0.1em}
\subsection{Experimental Settings}
\vspace{-0.2em}
In our experiment, the VVC reference software VTM-12.0 is used as the baseline. The codec adopts the Low-delay P (LDP) configuration according to the VVC Common Test Condition (CTC). The test sequences from classes B to E with different resolutions are tested as specified in \cite{CTCdocument}. For each test sequence, quantization parameter (QP) values are set to 22, 27, 32, 37, and Bjontegaard Delta-rate (BD-rate) \cite{2001Calculation} is used as an objective metric to evaluate coding performance. 

\vspace{-0.2em}
\subsection{Performance}
\subsubsection{Overall Performance Under Common Test Conditions} The R-D performance of the entire UAMM-integrated inter prediction framework on the VVC common test sequences is illustrated in Table~\ref{tab:CTC}. Y, U, and V represent the R-D performance gain of the three channels of YUV, EncT and DecT represent the encoding/decoding time. We can see that our proposed UAMM can achieve, on average, 0.13\%, and achieve up to 0.38\% BD-rate reduction (Y component) on VTM-12.0
for all sequences under the LDP configuration, which does not obviously increase the time complexity in the encoding/decoding process. The experimental results show that the proposed framework performs better for sequences with moving objects, such as $BasketballDrill$, $RaceHorses$, $Cactus$, and $Marketplace$.

\subsubsection{Comparison with the Other Temporal Motion Modeling Methods}
In VVC reference software (VTM-12.0), the technology of inter prediction most relevant to temporal motion modeling is the \textit{temporal motion vector prediction (TMVP)} \cite{McCann2013} and \textit{sub-block-based temporal motion vector prediction (SbtMvp)}\cite{yang2021subblock}. Both our method and these methods aim to achieve a flexible estimation of the motion field and match the appropriate motion trajectory in the temporal domain. In Table~\ref{tab:CTC}\,, \textit{TMVP} and \textit{SbtMvp} are all enabled in the VTM anchor by default, under the LDP configuration. In comparison with anchor, UAMM-integrated inter modes achieve good performance with \textit{TMVP} and \textit{SbtMvp} enabled. It can be concluded that UAMM can efficiently deal with some complex motion scenarios that \textit{TMVP} and \textit{SbtMvp} are difficult to address. 

\begin{table}
	\renewcommand\arraystretch{1.2}
	\centering  
	\vspace{-2em}
	\caption{BD-rate Results (Y Component) of Our Proposed UAMM Compared to VTM-12.0 on Specific Scenes}
	\vspace{-0.7em}
	\label{tab:Spscene}
	\setlength{\tabcolsep}{0.7mm}
	{
	\begin{tabular}{ccccc}
	\hline
	\textbf{Sequence}    &                      & \textbf{Resolution} &                      & \textbf{LDP (\%)} \\ \cline{1-1} \cline{3-3} \cline{5-5} 
	\textit{Beauty}  &                    & \textit{1080P}       &                      & -0.16\%               \\ \cline{1-1} \cline{3-3} \cline{5-5} 
	\textit{Jockey}            &                      & \textit{1080P}        &                      & -0.19\%               \\ \cline{1-1} \cline{3-3} \cline{5-5} 
	\textit{WestLibrary}            &                      & \textit{1080P}                &                      & -0.15\%               \\ \cline{1-1} \cline{3-3} \cline{5-5} 
	\textit{WestLabBuilding}            &                      & \textit{1080P}              &                      & -0.21\%               \\ \cline{1-1} \cline{3-3} \cline{5-5} 
	\textit{CrossRoad}            &                      & \textit{1080P}               &                      & -0.14\%               \\ \cline{1-1} \cline{3-3} \cline{5-5} 
	\textit{CrossRoad2}            &                      & \textit{1080P}               &                      & -0.12\%               \\ \cline{1-1} \cline{3-3} \cline{5-5} 
	\textbf{Overall}     &                      &  \multicolumn{3}{c}{\textbf{-0.16\%}}             \\ \hline
	EncT     &                                          & \multicolumn{3}{c}{101\%}             \\ \hline
	DecT     &                                          & \multicolumn{3}{c}{103\%}             \\ \hline

\end{tabular}
	}
	\vspace{-1.2em}
\end{table}

\subsubsection{Performance on the Specific Scenes}
According to the performance of UAMM under CTC, the proposed UAMM is beneficial to the scene in that the foreground and background are clearly separated, such as the surveillance-related scene ($Johnny$, $BasketballDrill$, $FourPeople$). To further verify the characteristics and advantages of UAMM, we test UAMM on some selected test sequences of specific scenes from some surveillance-related datasets \cite{xiph, yang2021spatiotemporal}. Table~\ref{tab:Spscene} shows the performance of these specific scenes. The results demonstrate the stable generalization of UAMM on specific scenes.

\begin{table}
	\renewcommand\arraystretch{1.3}
	\centering
	\fontsize{7.7pt}{9pt}\selectfont
	\caption{BD-rate Results of Some Sequences under LDB Configuration}
	\label{tab:LDB}
	\vspace{-0.92em}
	\setlength{\tabcolsep}{1.2mm}
	{
		\begin{tabular}{c|ccccc}
			\hline
			\textit{\textbf{Sequence}}         & \textit{BQMall}     & \textit{BQSquare} & \textit{Johnny} & \textit{BlowingBubbles} & \textit{BasketballPass} \\ \hline
			\textit{\textbf{BD-rate}} & 
			-0.19\%    & -0.35\%     & -0.11\%         & -0.11\%         & -0.15\%          \\ \hline
		\end{tabular}
		\vspace{-1.9em}
	}
\end{table}

\subsubsection{Bi-directional Prediction Exploration}
To further explore the potential of proposed method, we also test our proposed method on the Low-delay B (LDB) configuration, as shown in Table~\ref{tab:LDB}. The results of some sequences verify the potential of the proposed method in bi-directional setting.

\subsubsection{Limitation}
From the CTC results (Table~\ref{tab:CTC}), we find that the proposed method is limited to some motion scenes, such as the $BQTerrace$ (tiny objects with camera motion), $RitualDance$ (indistinguishability of foreground and background). Although the proposed acceleration model can handle some complex temporal motion fields to some extent, the characterization of extreme time motion remains a huge challenge in the way of the predefined motion model.

\section{Conclusion}
In this paper, the uniformly accelerated motion model is proposed to optimize the capacity of temporal motion modeling in VVC. Based on the theory of UAMM, the motion-related parameters (velocity, acceleration) are mined to assist the accurate ME and MC in the temporal domain. The experimental results demonstrate it can achieve good performance with a slight increase of time complexity compared to the VTM-12.0 anchor. For future work, first, we will extend the proposed motion model to improve more inter modes, such as the combination with the more flexible motion estimation schemes\cite{li2017efficient, li2015affine, li2022global, li2024object}, etc. Second, we will further explore the variable accelerated motion model to achieve more flexible motion modeling.

\bibliographystyle{IEEEtran}
\bibliography{IEEEexample}
\end{document}